# Partition test and sexual motivation in male mice

Kudryavtseva N.N.

Neurogenetics of Social Behavior Sector, Institute of Cytology and Genetics, Siberian Department of Russian Academy of Sciences, Novosibirsk, 630090, Russia,
natnik@bionet.nsc.ru

**Abstract**

Theoretical analysis of own and literature investigations of sexual motivation with the use of the partition test [Kudryavtseva, 1987, 1994] in male mice was carried out. It has been shown that appearance of a receptive female in the neighboring compartment of common cage separated by perforated transparent partition produces the enhancement of testosterone level in blood and stimulates the behavioral activity near partition as a reaction to the receptive female in naive males. In many studies this behavioral activity is considered as sexual motivation, arising in this experimental context in male mice. The lack of correlation between behavioral parameters and gonad reaction of males on receptive female, uninterconnected changes of these two parameters as well as the lack of sexual behavior between naive male and female when partition is removed cast doubt on this data interpretation. It has been supposed that in naive males behavioral reaction to a receptive female is induced by positive incentive – odor of the female associated with nursing and warmth from mother and other females which look after posterity. Short-term increase of the level of testosterone (possessing rewarding properties) is innate stimulus-response reaction which stimulates and prolongs behavioral interest of male to receptive female. It has been supposed that after sexual experience female odor is associated in experienced males with sexual behavior directed to the sexual partner and resulted in the formation of sexual motivation. The data are considered also in the light of the theory of motivated behavior including "liking", "wanting" and "learning" [Robinson and Berridge, 1993, 2000].

## Background

Since the 1950s influence of pheromones has been investigated in mice with the use of a small cage divided into two compartments by a partition (wire mesh) allowing the animals to see, hear and smell each other but preventing physical contact. It has been shown that a strange male located in the neighboring compartment can block incipient pregnancy of the female without direct physical contact [15, 17] by inducing changes in the hormonal background. In turn, female pheromones induce increase plasma testosterone and other hormones in male distantly [6, 7, 45], the intensity of enhancement is different in male mice of various strains [5, 46]. Neuroendocrine and neurochemical mechanisms of male sexual arousal induced by the presence of receptive female have been investigated [44, 45, 46, 48,]. Distant influence of males of different social status on the hormonal background of each other [15, 16] as well as behavioral reaction of males and females on each other depending on the stage of female estrus cycle [59] were also studied.

The similar cage with perforated transparent partition was used to form aggressive and submissive types of behavior in male mice in daily intermale confrontations [30]. As a result, the "partition test" [27, 29] was suggested as a tool for estimating behavioural reactivity of mice to the conspecific behind the partition dividing the experimental cage into equal parts. Number of approaches to the partition and total time spent near the partition, when mice touched the partition with their fore parts (nose, paws) were scored as indices of reacting to the partner in the neighboring compartment. The average time spent near the partition during one approach was evaluated by the ratio of the total time to the number of approaches to or touches of the partition. The behavioral response of animals to conspecific was shown to differ depending on physiological and psychological status of an individual, his social experience, the type of partner in the neighboring compartment and strain. Many experiments provided support in evidence of the fact that the "partition" test could be useful in experiments designed to study the mechanisms of sociability, anxiety, olfaction, aggressive and sexual behavior, cognition, psychoemotional disorders as well as in pharmacological studies of the above [31].

It became obvious that behavioral response to conspecific appears to reflect the motivation component of many behaviors based on previous experience or arising as instinctive response to innate specific (unconditional) stimulus (in this case pheromones) or partner behavior. The proof was obtained for anxiety and aggression. In male mice, behavioral parameters in the partition test were shown to correlate significantly with parameters in the elevated plus-maze test [36], which is sensitive to the action of anxiolytics [56]. It was shown that the less time a male spent responding to the partner in the neighboring compartment, the less time it spent in the open arms of the plus maze. This fact indicated that the partition test could be used to estimate level of anxiety in relevant experimental context. The same was found in the study of aggressive motivation: the more time a male spent near the partition responding to the other male behind it, the higher level of aggression this male exhibited in the subsequent encounter, when the partition was removed and animals could freely contact each other [28, 35]. In this case, as in other classical models of motivated behavior [42], the level of motivation was evaluated by the intensity of the corresponding consummatory act arising from the demonstration of intention (for example, aggressive behavior) or the absence of such (for example, in the case of anxiety and fear). It is these correlations that allowed using the partition test for the evaluation of the level of motivation in animals.

**Sexual motivation**

It was shown in many studies that a receptive female in the neighboring compartment of a common cage enhances the level of testosterone in sexually naive male [6, 46, 48]. This fact served as a base for the assumption that the test could be used for the measurement of sexual motivation. The first data proved that behavioral activity of males near the partition at presence of a receptive female was significantly higher than of a male [10, 34]. However it was known that receptive female is not the only driver of blood testosterone in males. Stress was shown to be accompanied by an increase in testosterone level in males too [22]. Notably, testosterone level increases proactively before aggressive confrontations [60].

As early as in the first experiments in male mice with different social status some doubts appeared that it is possible to interpret the partition behavior of males in the presence of receptive female in the context of sexual motivation. For example, aggressive males with the experience of victories in daily agonistic interactions were shown practically not to increase behavioral activity near the partition as a reaction to a female, which was put into the neighboring compartment immediately after the other male [10, 34]. At the same time, control and defeated animals doubtlessly displayed an interest in a receptive female. Preliminary data pointed to the absence of testosterone level increase in aggressive animals in response to the exhibition of a receptive female in contrast to the other two groups. This phenomenon was later explained by an inhibiting effect of chronic experience of aggression (forming an enhanced level of aggressive motivation) on sexual motivation or by an impaired sexual arousal in aggressive males. However the fact of the impairment of sexual motivation in aggressive males was paradoxical and contradicted extensive literature data indicating that it is the enhanced testosterone level that underlies the demonstration of aggression and reproductive success of dominant males [9, 40, 57]. But the second experiment on another mouse strain provided the verification of earlier data [32] on sexual motivation deficit in aggressive male mice.

Subsequently, however, after additional experiments the above results were interpreted not as impairment of sexual arousal *per se* but as impaired social recognition in aggressive males at social "male-female" and "male-male" interactions [13]. In experiment [3], with prior release from aggressive motivation, when the aggressive male was put in an individual cage for a day taken away from social confrontations, like control, aggressive males were found to display adequate behavioral response to the presence of a receptive female. Still, sexual motivation estimated by partition test in such males on the $30^{th}$ minute of measurement was reduced as compared with control, in which behavioral response to a female remained practically unchanged [3].

Obviously, to speak with certainty about changes in sexual motivation in each particular case it is necessary to obtain solid proof and conduct additional experiments. For example, it is necessary to find correlations between parameters of behavior near the partition in response to a receptive female prior to sexual interactions with certain forms of sexual behavior (sniffing, mountings, copulations, intromissions etc.) in the following direct contact with the sexual partner. Or a correlation is to be found between the parameters of behavior near the partition and some parameters of the gonadal hormone status, for instance, testosterone level. This would be a direct indication that the partition test could be used for measurement of the level of masculine sexual motivation in a particular experimental context.

However, the near-partition behavior in response toward a receptive female and hormone reaction turned out to be uncorrelated parameters. Moreover, with the partition removal sexual interaction between males and females failed to develop, at least, during the time of observation. A single study [3] finally detected a correlation between the parameters of male behavior near the partition and some behavioral patterns towards the female (mostly sniffing) after the removal of the partition. This study also failed to demonstrate pronounced sexual behavior. It is not clear whether naive or sexually experienced males were used in this study.

Analysis of literature data of the authors [4, 7, 50] who supposed that they were measuring sexual motivation by the partition test discriminating two components, behavioral and hormonal, also produced some doubtful moments:

1. In the above studies no significant correlations were found between behavioral parameters at the partition in naive males in response to a receptive female and blood testosterone level [4, 5];

2. Testosterone level increased to a maximum 20 minute after the introduction of the receptive female in the neighboring compartment and then (over half an hour) decreased to a control level. At the same time, behavioral reaction started immediately and remained at a high level after an hour [4, 50]. To some extent, this points to a lack of correlation between behavioral and hormonal components and to independent dynamics of these two characteristics in males;

3. In male mice of some strains a lack of both behavioral and hormonal reactions was found after 30 minutes exposure to BALB/c receptive female [5]. Earlier data obtained with the use of this experimental approach indicated that 40 minutes after the exhibition of a female in the neighboring compartment males not all strains reacted by a rise in testosterone level [46];

4. In male rats, the behavioral activity was enhanced pronouncedly with practically no change in testosterone level in response to a female put into the neighboring compartment [18];

5. Under effect of some serotonin preparations one component of sexual motivation (hormonal or behavioral) changed and the other remained unchanged, for example, pharmacological stimulation of 5-HT1B receptors reduced the time of behavioral response to a female and had no effect on testosterone level [4, 56];

6. Exposure to stress could also change one component of sexual motivation and produce no effect on the other. For example, restriction for five hours that took place a day before the test did not change the intensity of behavioral response to receptive female but significantly diminished testosterone level [4].

The reported data indicated that behavioral and hormonal components of the response to the presence of a receptive female in naive male might act as independent variables (rather than an indivisible reaction), which is likely to change according to its own laws. The direct indication that the term "sexual motivation" in the given experimental context bears a great amount of hypothesizing was the lack of demonstration of sexual behavior by naive males after the removal of the partition, when males and females could freely contact each other.

**Hypothesis.**

The effort to interpret the observations and answer the question "What is that the partition test actually measures in sexually naive males?" and also to offer, with regard for own and literature data, a logical explanation of what is taking place in the cage at distant interaction of naive males and receptive females has lead to a quite unexpected conclusion. In given experimental context, the partition test is most likely not to measure sexual motivation. To explain the absence of sexual interaction between mice of different sexes one is to assume that sexually naive males probably react to the odor that was positively rewarded in the babyhood – nursing and warmth provided by the mother and other females who looked after the progeny together. Males like mother's odor (sort of "Oedipus situation"), they spend a lot of time at the partition trying to get over with the reaction to the female lasting more than an hour. This is evidence of the appeal of female odor for males and positive emotions of the males. It is then reasonable to assume that for mouse strains whose males withheld reaction to receptive female of a other strain [5] the female odor was alien and too dissimilar to mother's odor so no behavioral reaction was exhibited. This also explains the absence of a correlation with testosterone level and the absence of sexual behavior.

However, if one agrees that in naive males exposed to the a receptive female the partition test does not measure sexual motivation, it has to be accepted that an increase of testosterone is independent of the behavior, arises instinctively as innate response to a species-specific incentive – the odor of sexual partner. It was shown experimentally that in naive males a behavioral reaction to the female and testosterone increase could be unrelated parameters. Alternatively, a short-term rise of testosterone level in response to a receptive female could be explained by a non-specific response to positive emotion, in which testosterone may be involved. Support for this premise could be found in the literature that shows the rewarding effects of testosterone [2, 19, 62]. However it was natural to assume that the behavioral and hormonal components of the response to a receptive female may become interconnected upon acquisition of sexual experience.

### Possible scenario of events in the given experimental context

In naive males, behavioral reaction (*sexual activation*) to the presence of a female arises immediately when the female is placed in the neighboring compartment of the cage. The reaction is induced by an incentive positively reinforced in the babyhood (odor of mother and other females). Additional component of the response could be the demonstration of communicative and exploratory activities toward a strange partner.

An increase in testosterone *(sexual arousal)* occurs in 20-40 min [4, 46, 50], which can be regarded as a delayed response to the female odor. This response is instinctive, based on stimulus-response behavior, where the stimulus—the female odor—is innate and species-specific. Then over half an hour the testosterone level decreases but the behavioral reaction lasts for a long time. A question arises: why so long? In the experiments focusing on the duration of the host male response to a strange male exhibited in the neighboring compartment of common cage behavioral activity at the partition as a response to the male decreased relatively quickly [26]. The prolonged behavioral reaction to the female is likely to be rewarded by the enhanced level of testosterone, which is capable of inducing positive emotional reaction all by itself. It may be assumed that even a short rise of testosterone level tints a distant contact with a female with positive emotions, creating positive reinforcement even distantly.

With maturity accompanied by an increase of testosterone level at the encounter with female odor as a source of sexual arousal there arise associative relations that form goal-directed behavior of female search aimed at obtaining positive emotions from enhanced blood testosterone level. Prolonged sexual excitation without coping could lead to the development of a sexual pathology.

Once the first sexual experience occurred, the male is aware of the positive implications of sexual catharsis that brings pleasure and relief over satisfaction. These positive implications build up *sexual motivation*, which is conceived as a behavior directed to satisfying the existing desire. Thus, motivation, which forms the base of goal-directed behavior, comes to existence only with experience [58]. The use of the term "sexual motivation" for the description of naive male behavior in response to a female appears to be somewhat misguiding in the frame of this theory as naive males are lacking sexual experience.

Preliminary investigation of experienced and naive males in the experiment under consideration showed that the overwhelming majority of experienced males (who participated in reproduction and had sexual successes) after a 5-days refractory period at room lighting from the beginning demonstrated highly pronounced behavioral activity at the partition in response to a receptive female and then a expressed sexual behavior (anogenital sniffing, mountings and copulations) during 20 minutes of observation from the moment of partition removal. In naive males (who demonstrate weaker reaction to the female) after partition removal aggressive attacks on the female were observed as well as considerably reduced sniffing activity compared with experienced males. Some of the naive males made mounting attempts but after 10 minutes both males and females lost interest in each other. In general, it can be stated that naive males do have interest in the female while lacking sexual motivation (as judged by the missing consummatory act) or having such at a significantly lower level than sexually experienced males. At the beginning it was difficult even to decide whether it was sexual interest to the female or just interest in a new partner, in which anogenital sniffing is also present. The main conclusion made on the basis of these observations was the necessity to use a sexually experienced female as an incentive rather than a naive female since the behavior of the latter, who is unaware of what the naive male is wanting from her, may produce a inhibiting effect on the demonstration of his subsequent sexual behavior in naive male mice.

**Liking, wanting, learning.**

Motivation theories advanced in different years were reviewed extensively by K.S. Berridge [12]. Jointly with T.E. Robinson he suggested a biopsychological theory of addictions, which was further elaborated in later years [54, 55]. In the study of psychological and neurobiological prerequisites of drug addictions the authors separated two motivation components: "liking" (getting pleasure from drugs) and "wanting" (drug abuse, motivation to take the drugs). It was suggested that the two components may have different neural regulation. "Liking" is regulated by the brain's opioidergic and GABA-benzodiazepine systems while the regulation of "wanting" involves dopaminergic mediation paths with different brain structures being involved in the processes. Both "wanting" and "liking" may operate outside of subject's conscious awareness [11]. As the drug addiction develops (with experience acquired - learning) the balance between the two components may shift to "wanting" [54, 55].

This conception was extended to food motivation – food consumption is viewed as positive reinforcement, in which two independent components can also be separated: "liking" – pleasure from the food (hedonistic component) and "wanting" – proper food motivation [11, 21, 41], which are supposed to have different mechanisms. For instance, we are ready to eat the food that we do not like when we are hungry. Or we are not hungry but will eat the food because we like it.

The components of motivational behavior "liking", "wanting" and "learning" appear to be applicable for the explanation of the mechanisms of sexual motivation. The role of learning (or experience) in sexual behavior has been studied substantially [49]. Sexual behavior is commonly understood as the behavior aimed at getting the pleasure from copulation, the notion of which individuals gain by different means from the initial experience. Learning (sexual experience) is a key to all further forms of sexual behavior such as search, preference for certain partner, obstacle overcoming, courtship, sexual arousal conditioned by perceived positive reinforcement, and copulation [49]. It may be assumed that repeated sexual experience is forming the motivation directed to both getting pleasure ("liking") and satisfying wanting in case of prolonged refractory period.

In sexually naive males the behavioral reaction to a receptive female involves "liking" and at first is not associated with sexual behavior. "Liking" is driven by female odor and positive emotions from enhanced testosterone level in her presence. Sexual experience accompanied by powerful positive reinforcement on the level of physiological reactions resulted from sexual interactions leads to the formation of motivation for its repeated occurrence, which turns into the need (wanting). B. Everitt [20] has presented evidence of different mechanisms of sexual arousal and its realization (intromissions and ejaculation).

It may be assumed that the two components of sexual motivated behavior, "liking" and "wanting", could manifest themselves in association as they are associated forms of the same state, but at the same time they can exist all by themselves, independently of each other. For example, in naive males there is only "liking". In sexually experienced males there are both "wanting" and "liking", but "liking" implicates sexual contact. It may be anticipated that these males are most likely to show a correlation between behavioral activity at the partition and testosterone level. It is expected that correlations with sexual behavior should be found in males with the matured sexual drive and a sufficiently long refractory period after the last copulation.

Studies of male behavior in response to a female depending on the presence or absence of sexual experience of the male produced implicit evidence in support of this hypothesis. Naive male rats having to make a choice were shown to prefer an incentive from the female but this preference was reduced as compared with what was demonstrated by experienced males. A similar result was obtained in the test when a male could see, hear and smell an experienced male or a receptive female without contacting them physically [53]. It was shown that it took sexually experienced males less time to run up to a receptive or nonreceptive female [39] and to make a correct choice of an estrus female in the T-maze as compared with naive males [40]. Some authors considered the behavior of a naive male toward a receptive female as unconditioned sexual incentive motivation, while the reaction of a sexually experienced male – as conditioned sexual incentive motivation [1, 49].

**The features of the partition test application in the study of sexual motivation in male mice**

*Adequate partner choice.* In the studies of sexual motivation in mice of different strains, strange as it may seem, it is most difficult to choose an adequate sexual partner – the female. Such studies [5, 46] commonly used a receptive female of BALB/c strain, which was supposed to have an attractive and powerful odor for male mice. However, males of not all strains reacted to BALB/c female by the enhancement of testosterone level and behavioral activity in the partition test. The obvious conclusion is that sexual motivation in these males is diminished. But it is well known that mice of inbred strains are distinct as regards many psychophysiological characteristics: emotionality, locomotor and exploratory activities, anxiety, sensitivity to olfactory incentives and ability to discriminate such incentives etc. Any of these features may produce an effect on the response intensity under exposure to a strange receptive female placed in the neighboring compartment of the cage. For instance, it may be assumed that some males do and other males do not feel the smell, which may have nothing to do with the mechanisms of sexual motivation as such. Back in 1988 S. N. Novikov in his excellent monograph "Pheromones and reproduction of mammals" [47] analyzed data on testosterone level increase in male mice of different strains in the presence of a receptive female [46] and raised the questions, the convincing answers to which have not been found yet: "What is the relative role of olfaction in the regulation of this effect in different strains?", "Is the female genotype of importance for the manifestation of sexual activation effect?" and "What neurophysiological mechanisms underlie low reactivity of males of some strains?" It is doubtful that males of the strains that did not react to the BALB/c females breed worse than males of other strains or that sexual motivation of such males is reduced. This example illustrates the complexity of the task of sexual partner selection for the experiments, which is to be further elaborated. It is logical to suppose that if males were offered a female of the same strain, the behavioral and hormonal reaction would be found. Without excluding a standard tester, to avoid multiple interpretations in such experiments it is recommended to investigate also the reaction of males to a receptive female of the same strain, whose most psychophysiological parameters will be the same as those of the male. If males react to a receptive female of the same strain, this will explain the absence of reaction to a receptive female of another strain.

*Dynamic changes of testosterone level.* Different hormonal response in males to the presence of a receptive female could be due to different dynamics of blood testosterone changes conditioned by a great number of physiological mechanisms. In males of some strains testosterone level may rise twenty minutes (or earlier) after the exhibition of a receptive female [4, 5] and in half an hour not differ from the level in control males. In males of other strains this process may begin later and testosterone enhancement is observed after 40 minutes of exposure [46]. This is why detailed investigations are needed to obtain a dynamic picture of testosterone level changes to make an adequate conclusion on high or low reactivity of the males of a particular strain to the odor and presence of a receptive female.

*Effects of pharmacological treatment and stress.* Obviously, any pharmacological preparation under system administration can influence many forms of animal behavior and physiological status. This refers in particular to those preparations that change psychoemotional state, say, by activating or inhibiting neurotransmitter systems of the brain. In such cases it is very difficult to identify specific pharmacological effect on this or that form of behavior. In the studies of sexual motivation it was shown that stimulation and blocking of serotonin receptors of different types [8, 51, 52] most frequently led to a reduction of both the hormonal and behavioral components of the response to a receptive female (although to a different degree). One-way change of the parameters studied could be evoked by changes in the activity of serotonergic system, which is commonly associated with, for example, the development of anxiety [56], which is capable as such of suppressing the behavioral response and testosterone level of the male at the exhibition of a receptive female. That is why such experiments fail to provide unequivocal explanation of what actually happened: whether anxiety (fear) has grown or the sexual motivation has lowered. An increase of anxiety could be a reason of the changes of behavioral and hormonal response of the male under chronic stress, which is probably observable at daily restriction in prenatal age [38]. Additional experiments are needed to understand the specific character of changes in sexual motivation or this phenomenon has to be explained with the use of literature data.

***The features of interpretation of behavioral parameters in the partition test.*** Various approaches to the study, possible experimental designs with the use of the partition test are described in the review article published by the author [31]. However, new data have been accumulating on the use of the test in other laboratories, which seems not always to be interpreted correctly. The partition test is actually simple, but the main difficulty is not in measuring the number of times a male "sticks his nose and forepaws" into the partition [50] but to interpret behavioral parameters. It is tempting to suggest that the number of approaches reflects motor activity and the duration of stay at the partition – the level of some motivation. However, the number of approaches, obviously, might reflect motor activity only when the neighboring compartment is vacant and clean and the male under investigation has explored the territory of his compartment for not less than a day and passed the period of activation before the test. After the appearance of a partner behind the partition there arises a motivational component of behavior so the assertion that the number of approaches reflects motor activity is to be verified in the tests expressly measuring motor and exploratory activity. Sometimes a clue to understanding is in the derivative parameter of average time spent near the partition on one approach.

Let us imagine the following situation. It was shown that under exposure (for instance, administered preparations) the number of approaches to the partition decreases with the total time of reaction to a partner in the neighboring compartment not changing significantly. If the study deals with the response of male to receptive female the result could be interpreted by a reduced motor activity without influence on the motivation. However if you calculate the average time of one approach by dividing the total time spent near the partition into the number of approaches, it may increase as compared with control males. This means that a male approaches the partition and does not retreat from it craving for the female, while moving continuously at the partition without being distracted for walks around the cage. This points not to a decrease in motor activity but to the enhancement of sexual motivation. One more example is when the number of approaches increases while the average time per one approach significantly decreases as compared with control males. In this case you would rather suggest not an increase in motor activity but the development of two opposite motivations. A mouse first comes to the partition then retreats from it. This "approach-avoidance behavior" is driven by the urge to explore a new object and by fear before the unknown [43]. In psychiatry this phenomenon, if it occurs in humans in an acute form, is called ambivalence, meaning that the same incentive is attractive and aversive at the same time. A significant decrease of the average time spent near the partition per one approach was observed in our study of depressive mice [33]. Consideration of only two parameters of the "partition" test narrows the scope of possible interpretations and, thus, the understanding of the observed phenomena.

**Conclusion: Experimental approaches to the standardization of experiments in the study of mechanisms of social interactions.**

The point here seems to be about the obvious thing, standardization of experiments. It is generally accepted that in the experiments the situation for animals is to be maximally equalized, i.e. conditions for control and experimental animals should be the same, excluding the incentive the response to which is being studied. For example, in the experiments measuring the response of animals to a pharmacological treatment control and experimental animals are administered a vehicle and drug, respectively. If animals with different physiological status are investigated, the rule is to equalize the incentive the response to which is being studied. Let it be called a conventional approach.

However common rules of experiment standardization go to pieces when one speaks about social behavior, which is understood as any interaction of two or more individuals on any occasion. Twenty years ago a fact was first confronted (verified later many times) that one and the same stimulus influences differently the behavior and physiological state of animals with different social status. If a finger is brought to the nose of a male with the experience of aggression he may bite, a submissive individual with the experience of social defeats (victim) may run away and an intact animal - exhibit exploratory behavior by sniffing it. This means that motivational and physiological mechanisms of the above behavioral patterns are different and that in individuals with different social status the same stimulus results in entirely different consequences. Similar results were obtained in the study of the influence of different forms of psychoemotional stress on males and females in the work by D. F. Avgustinovich and I. L Kovalenko: social stimuli that induced severe stress in males causing the development of psychoemotional disorders produced no effect on females [25]. The same happens in human life – what is important for one individual is insignificant for another. As for the subject matter, one and the same incentive, receptive female, may activate totally different motivational components of behavior in naive and experienced males by inducing totally different physiological and behavioral mechanisms.

Therefore, in some experiments it is necessary to search specific incentives (sometimes they are different) producing similar effect on individuals with different emotional status or on male and female, thereby equalizing not the incentive as such but the condition it induces. This means that incentives might be different but their effect - similar. Obviously, to induce sexual motivation in males of different strains you need an appropriate incentive – odor or presence of female. For humans, an appropriate incentive is, in the first place, a visual image of sexual partner. In both cases sexual motivation arises following the same physiological rules. Let us call the approach of using an adequate incentive "adequate", which, as shown by experience, is preferable for the investigation of social behavior. Moreover, it is to be noted that taking an adequate approach may require not so much the use of different incentives for inducing a similar response as totally different conditions to find these or those physiological and psychoemotional peculiarities of each particular individual. For example, in the studies of sexual arousal mechanisms in males with repeated experience of aggression they are to be placed in the conditions that release them from aggressive motivation [3], after which they are capable of reacting to a receptive female adequately. In animals with repeated experience of social defeats receptive female might evoke a behavior at the partition similar to control [32] but it is insufficient to conclude that sexual behavior of such males is intact since social stress is known to lead to the development of gonadal hypofunction [37] and reproduction impairment [24]. Additional experiments are needed to say whether the stress of social confrontations affects sexual arousal and, if so, how the affected sexual arousal influences sexual behavior as such. For this purpose detailed investigation of sexual behavior is needed.

Summing up, it may be concluded that in the studies of social behavior for adequate interpretation of the behavioral phenomena are a properly elaborated experimental context and the understanding of physiological regulation of the phenomenon under study, which allows unequivocal interpretation of the motivational component of a behavioral pattern.